\documentclass[%
reprint,
superscriptaddress,
amsmath,amssymb,
aps,
]{revtex4-1}

\usepackage{graphicx}
\usepackage{bm}
\usepackage[dvipsnames]{xcolor}
\usepackage{tabularx}
\newcolumntype{Y}{>{\centering\arraybackslash}X}

\newcommand{\efield}{{\mathcal{E}}}

\newcommand{\acos}{\mathrm{acos}}
\newcommand{\quemener}{{Qu\'em\'ener}\ }

\begin{document}

\title{Microwave shielding with far-from-circular polarization}
\author{Tijs Karman}
\email{tijs.karman@cfa.harvard.edu}
\affiliation{ITAMP, Harvard-Smithsonian Center for Astrophysics, Cambridge, Massachusetts 02138, USA}
\date{\today}
\begin{abstract}
Ultracold polar molecules can be shielded from fast collisional losses using microwaves,
but achieving the required polarization purity is technically challenging.
Here, we propose a scheme for shielding using microwaves with polarization that is far from circular.
The setup relies on a modest static electric field,
and is robust against imperfections in its orientation.
\end{abstract}
\maketitle

Ultracold polar molecules are emerging as a platform for quantum science and technology with applications in precision measurement \cite{v2018improved}, quantum simulation \cite{Santos:2000,Micheli:2006,Baranov:2012}, and computing \cite{DeMille2002,Yelin2006,ni2018dipolar}.
Many species of ultracold molecules are now realized experimentally at ultracold temperatures,
either by associating ultracold atoms \cite{Ni:KRb:2008,Takekoshi:RbCs:2014, Molony:RbCs:2014, Park:NaK:2015, Guo:NaRb:2016,Rvachov2017,yang2019singlet},
or by directly cooling molecules \cite{Truppe:MOT:2017, McCarron:2018}.
The lifetime of ultracold molecules is limited by collisional losses \cite{Ye2018,Guo2018,Gregory2019},
even at a typical molecular density of {$10^{-10}$~cm$^{-3}$} that is orders of magnitude below that required for some applications.
The collisional loss rates observed are on the order of the universal loss rate \cite{Idziaszek2010},
suggesting the loss occurs at short range when the molecules approach one another closely.
For some molecules this short-range loss is attributed to two-body chemical reactions \cite{Ospelkaus2010}.
Nonreactive molecules undergo effective two-body loss at much the same rate,
presumably mediated by the formation of long-lived collision complexes \cite{Mayle2012},
that may subsequently be lost through three-body recombination \cite{Christianen2019b} or photoinduced processes \cite{Christianen2019}.

Collisional losses can be suppressed generally by inducing long-ranged repulsive interactions that prevent the molecules from reaching short range.
This is referred to as shielding.
\quemener and Bohn have suggested electrostatic shielding of polar molecules in the $n=1$ rotationally excited state \cite{quemener2016shielding}.
This may require strong electric fields in the order of $3.25 b/\mu$ \cite{gonzalez2017adimensional},
where $b$ is the rotational constant and $\mu$ the dipole moment.
This requirement may be circumvented by microwave shielding \cite{Karman2018a,lassabliere:18},
where ground state molecules can be shielded by inducing repulsive resonant dipole-dipole interactions through microwave dressing with $n=1$ rotationally excited states.
Furthermore, Gorshkov \emph{et al.}\ suggested using combined static and microwave fields to achieve shielding by a repulsive second-order interaction, after precisely canceling the first-order interactions due to both fields \cite{Gorshkov2008}.
While microwave shielding is feasible in static fields,
the mechanism was later shown to be different and not reliant on precise cancellation of first-order interactions \cite{Karman2018a}.

The technically challenging requirement for implementing microwave shielding is realizing almost pure circular polarization,
especially in the presence of reflections of microwaves off the vacuum chamber.
Microwave shielding is effective for circularly polarized microwaves \cite{Karman2018a},
but not for linear polarization.
The reason is the coupling to the repulsive branch of the resonant dipole-dipole interaction, which provides shielding, depends on the orientation of polarization relative to the intermolecular axis \cite{Karman2019a}.
In the case of linear polarization, collisions along the polarization direction are not shielded,
while nonadiabatic transitions to lower field-dressed states lead to rapid loss.
For circular polarization, collisions along all directions are shielded and nonadiabatic transitions are suppressed for sufficiently high Rabi frequencies.
For elliptical polarization, nonadiabatic transitions cannot be fully suppressed,
and effective shielding requires a polarization that is 90~\% circular in the field \cite{Karman2019a},
or equivalently has a 20 dB power extinction ratio between $\sigma^+$ and $\sigma^-$ components.

\begin{figure}[h!]
\begin{center}
\includegraphics[width=0.475\textwidth]{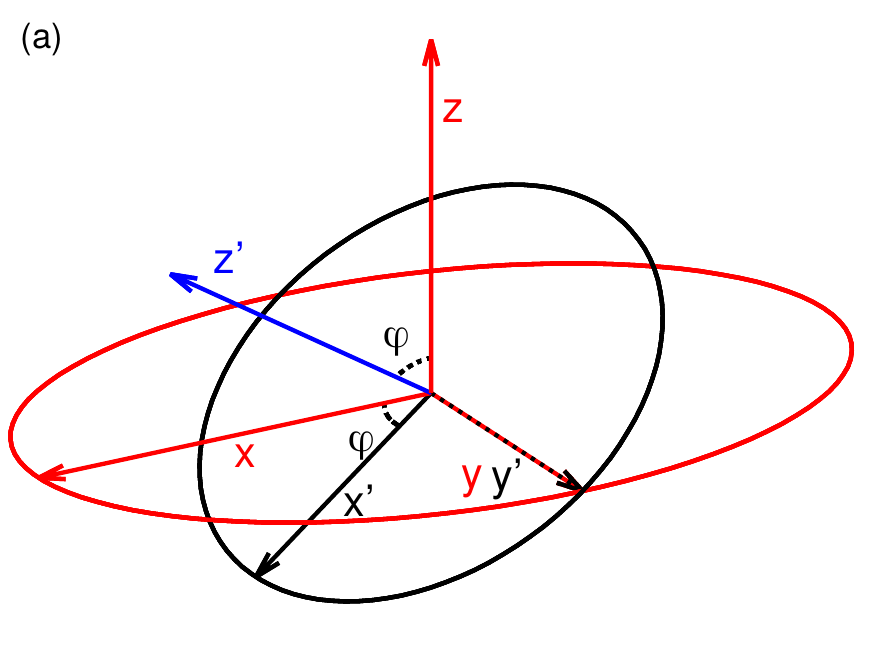}
\includegraphics[width=0.475\textwidth]{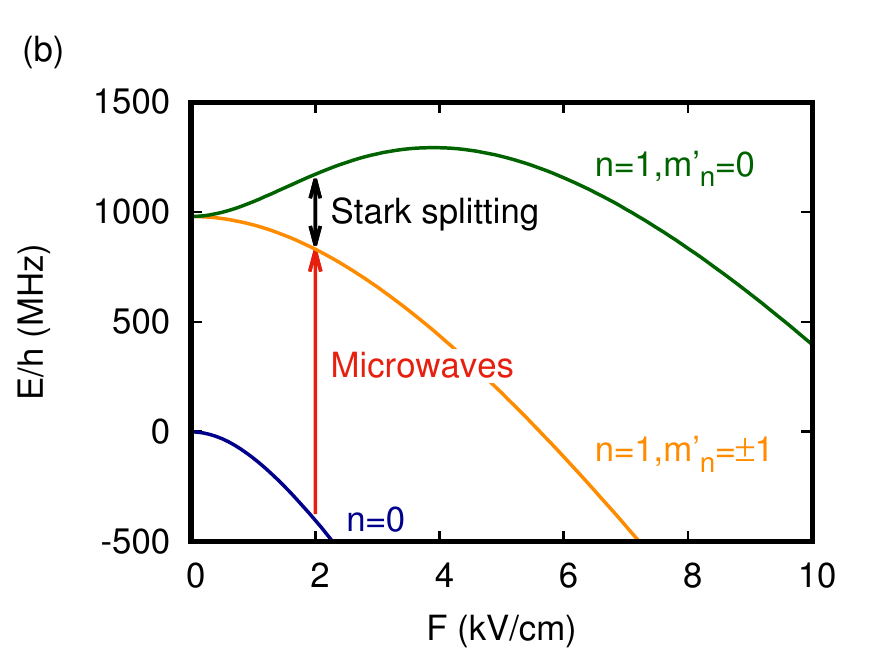}
\caption{ \label{fig:polellipse}
(a)
Red shows a general polarization ellipse where the $x$ and $y$ axes are chosen as the semi-major and semi-minor axes.
This is is equivalent to circular polarization in the $x'y'$ plane, shown in black, plus an additional linear polarization component along $z'$, shown in blue.
The coordinate transformation is discussed in the text.
(b)
A static $\efield$-field along $z'$ lifts the degeneracy of $m_n'=0$ and $m_n'=\pm 1$ states,
such that the $n=1,m_n'=0$ state is Stark shifted out of resonance and the linear polarization component along $z'$ has no effect.
Under these conditions, the Hamiltonian reduces to that of polar molecules in the presence of circularly polarized microwaves and a static $\efield$-field,
which has previously been shown to realize effective shielding \cite{Karman2018a}.
}
\end{center}
\end{figure}

In this work, we propose a modified scheme for microwave shielding that is effective for polarizations that are far from circular.
The main idea is illustrated in Fig.~\ref{fig:polellipse}.
General elliptical polarization can be characterized as
\begin{align}
\sigma(\xi) &= \sigma_+ \cos\xi - \sigma_- \sin\xi \nonumber \\
&= -[\sigma_x \sin(\xi+\pi/4) + i \sigma_y \cos(\xi+\pi/4)],
\end{align}
which interpolates between $\sigma_+$ circular polarization at $\xi=0$ and $\sigma_x$ linear polarization at $\xi=\pi/4$.
For $\xi$ in this range, the semi-major and semi-minor axes are along $x$ and $y$.
We then consider a transformation to a coordinate frame, $x'y'z'$, defined by a rotation about the semi-minor axis, $y$, by an angle
\begin{align}
\varphi=\acos[\cot(\xi+\pi/4)]
\label{eq:phi}
\end{align}
 such that the polarization becomes {$\sigma_{+'} \cos(\xi+\pi/4) \sqrt{2} + \sigma_{z'} \sin(\varphi)\sin(\xi+\pi/4)$}.
That is, any polarization can be thought of as perfectly circular in the $x'y'$ plane plus an additional linear component along $z'$.
Next, we consider applying a static $\efield$-field along $z'$,
which lifts the degeneracy of $m_n'=0$ and $|m_n'|=1$ states,
where $m_n'$ is the $z'$ projection of the rotational angular momentum $\bm{n}$.
This Stark shift serves to shift the $m_n'=0$ state, addressed by the spurious $\sigma_{z'}$ polarization component, out of resonance while the microwaves are kept tuned to the $m_n'=\pm 1$ states.
Under these conditions, this Hamiltonian effectively reduces to dressing with microwaves that are purely circularly polarized about a static external field,
which has previously been demonstrated to realize effective shielding \cite{Karman2018a}.

We theoretically treat molecule-molecule collisions as in Refs.~\cite{Karman2018a,Karman2019a}, and briefly summarize the approach here:
The molecules are treated as rigid rotors that interact with one another through the dipole-dipole interaction,
and with static and ac electric fields through the Stark interaction.
We then perform coupled-channels scattering calculations where at long range we match to the usual scattering boundary conditions,
and at short range we match to a completely absorbing boundary condition \cite{janssen:13},
in the spirit of the universal loss model \cite{Idziaszek2010}.
The short-range boundary condition is imposed at $R=20~a_0$,
and the channel basis is truncated at $n_\mathrm{max}=3$ and $L_\mathrm{max}=6$.
We distinguish between losses due to reaching short range (RSR),
and microwave-induced loss (MIL).
The latter corresponds to inelastic scattering into lower-lying field-dressed levels,
which is referred to as MIL as these channels are not present in the absence of microwave radiation.
The role of hyperfine degrees of freedom was addressed previously \cite{Karman2018a,Karman2019a},
and hyperfine is not included here as its effects can be suppressed by applying a modest magnetic field.

\begin{figure}[t!]
\begin{center}
\includegraphics[width=0.475\textwidth]{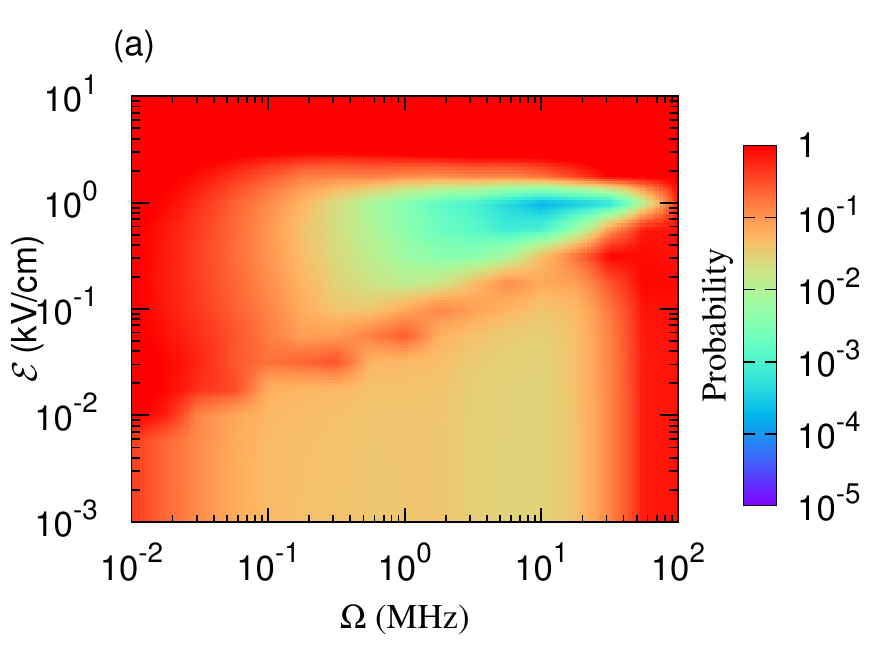}
\includegraphics[width=0.475\textwidth]{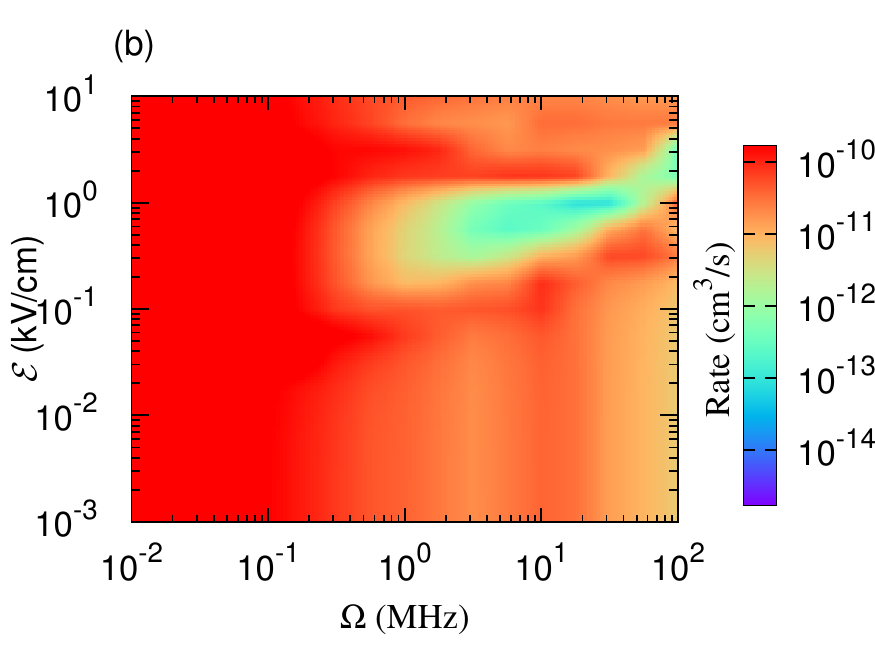}
\caption{ \label{fig:OmegaF}
Probability of RSR (a) and MIL rate (b) as a function of $\Omega$ and static field strength $\efield$ for $\xi=\pi/8$, \emph{i.e.}, half way between $\sigma^+$ and linear $\sigma_x$ polarization.
For high enough $\Omega$ and intermediate field strengths, losses due to RSR and MIL are simultaneously small and good shielding is obtained.
A larger Stark splitting is required for increasing Rabi frequency,
resulting in the triangle-shaped region of effective shielding.
}
\end{center}
\end{figure}

Figure~\ref{fig:OmegaF} shows the probability of RSR and MIL rate for RbCs molecules at 1~$\mu$K as a function of the Rabi frequency, $\Omega$, and static field strength, $\efield$.
This is obtained on resonance, $\Delta=0$, and for fixed ellipticity, $\xi=\pi/8$,
which is half way between $\sigma_+$ circular polarization at $\xi=0$ and $\sigma_x$ linear polarization at $\xi=\pi/4$.
The static field is applied along the $z'$ axis,
\emph{i.e.}, between the polarization ellipse's normal and semi-major axis at an angle $\varphi$, see Eq.~\eqref{eq:phi}.
At low field strength, shielding is ineffective because the polarization is far from circular.
At high field strength,
dipolar scattering becomes dominant and leads to high loss rates.
At intermediate field strengths, there exists a region where losses due to RSR and MIL are small simultaneously, and effective shielding is realized.
Shielding at higher Rabi frequency, $\Omega$, requires larger Stark splittings and hence field strengths, $\efield$,
leading to the triangular shape of the shielded region in Fig.~\ref{fig:OmegaF}.
Shielding of losses to $2 \cdot 10^{-13}~$cm$^3$/s -- three orders of magnitude below the universal loss rate of $1.7 \cdot 10^{-10}~$cm$^3$/s -- is obtained for Rabi frequencies around $\Omega=10$~MHz and field strengths around $\efield=1$~kV/cm.

\begin{figure}
\begin{center}
\includegraphics[width=0.475\textwidth]{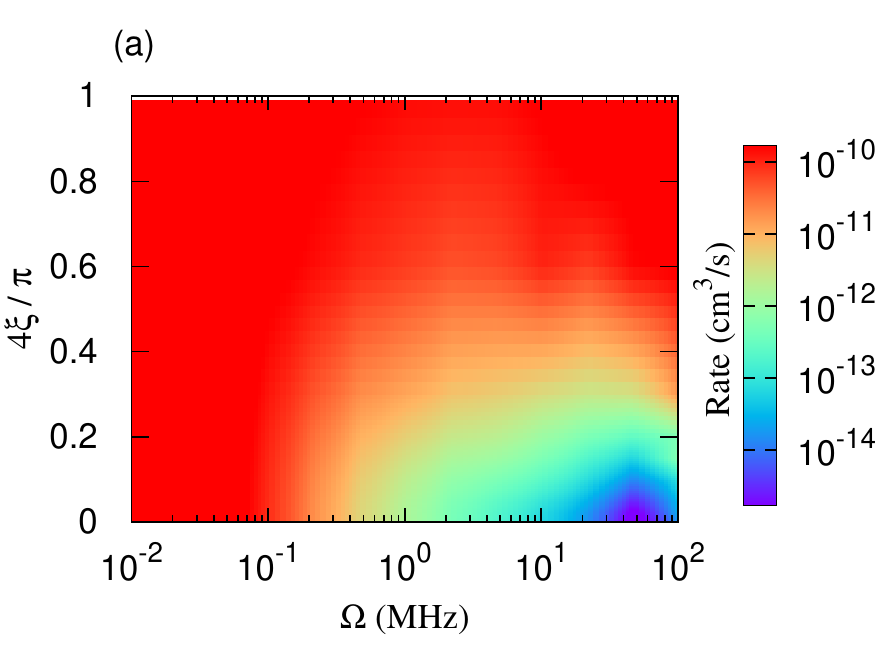}
\includegraphics[width=0.475\textwidth]{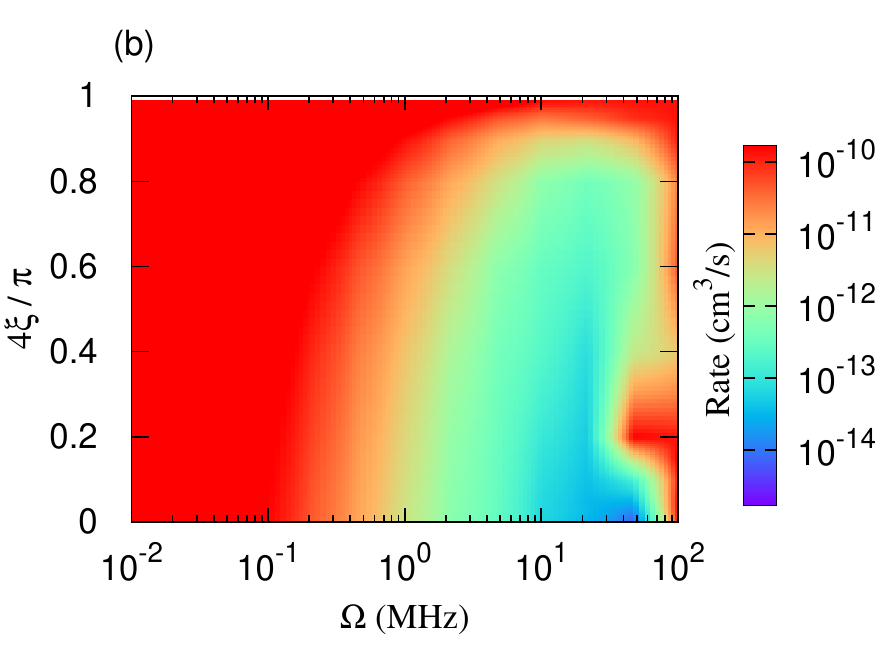}
\caption{ \label{fig:Omegaxi}
Total loss rate due to both RSR and MIL as a function of $\Omega$ and $\xi$ for $\efield=0$ (a) and $\efield=1$~kV/cm (b).
At zero field, shielding requires $4\xi/\pi \le 0.1$, \emph{i.e.}, an imperfection in the ellipticity angle of less than 10~\%.
Including a static field along $z'$ enables shielding by elliptically polarized microwaves with large eccentricity, up to $4\xi/\pi \approx 0.8$.
}
\end{center}
\end{figure}
\begin{figure}[ht!]
\begin{center}
\includegraphics[width=0.475\textwidth]{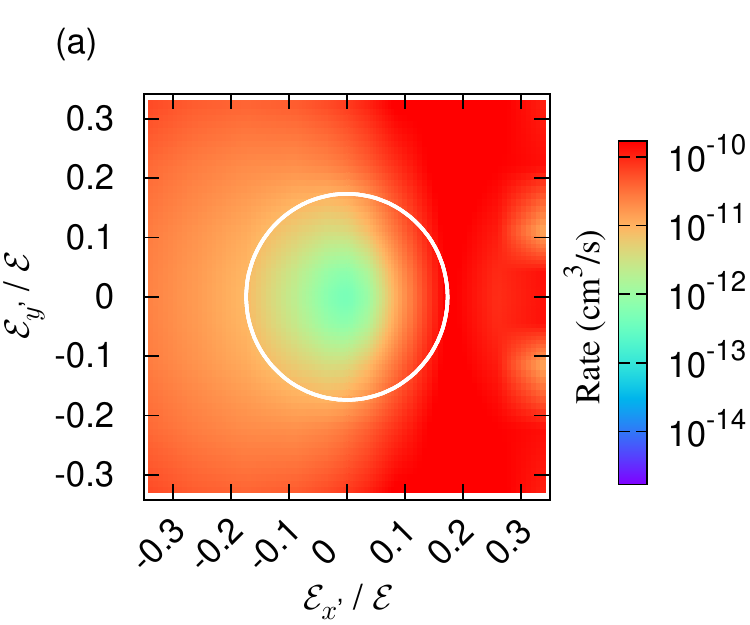}
\includegraphics[width=0.475\textwidth]{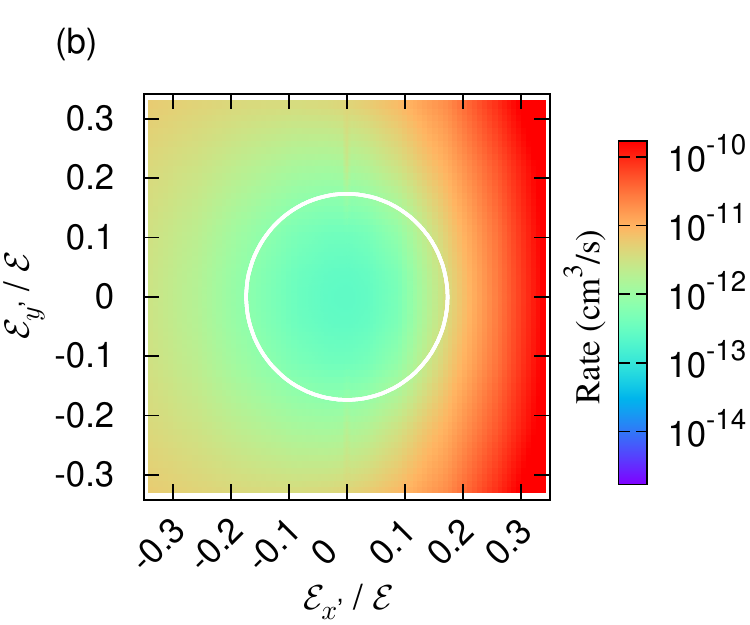}
\caption{ \label{fig:xy}
Total loss rate due to RSR and MIL as a function of $\efield_{x'}$ and $\efield_{y'}$ imperfections in the orientation of a $\efield=1$~kV/cm static field.
Panel (a) and (b) correspond to $4\xi/\pi=3/4$ and $4\xi/\pi=1/2$, respectively.
These polarizations can be thought of as 25~\% and 50~\% circular, respectively.
In both cases, the polarization ellipse is too eccentric to provide shielding without the additional static field.
The white circle indicates an imperfection in the orientation of the static field of 10$^\circ$.
Shielding of losses below $10^{-12}$~cm$^3$/s for $4\xi/\pi=3/4$ requires orientating $\efield$ along $z'$ to within 5$^\circ$,
and the tolerance is even more forgiving for polarizations that are closer to circular.
}
\end{center}
\end{figure}

Figure~\ref{fig:Omegaxi} shows the total loss rate, due to both RSR and MIL,
as a function of Rabi frequency, $\Omega$, and microwave polarization ellipticity, $\xi$.
Panels (a) and (b) correspond to static field strengths of $\efield=0$ and $\efield=1$~kV/cm, respectively.
The static field is applied along the $z'$ axis,
which again lies at an angle $\varphi$, see Eq.~\eqref{eq:phi}, between the polarization ellipse's normal and semi-major axis.
At zero static field, shown in panel (a), shielding is effective only for polarizations close to circular, $4\xi/\pi < 0.1$,
which corresponds to a power extinction ratio of the $\sigma_-$ and $\sigma_+$ microwave field components larger than 22~dB \cite{Karman2019a}.
For an achievable\cite{ni2009dipolar,ni2009quantum,Park2015,covey2018} field strength of $\efield=1$~kV/cm, shown in panel (b),
effective shielding can be obtained for polarizations that are far from circular,
with eccentricity up to $4\xi/\pi \approx 0.8$.

So far we have demonstrated excellent shielding can be recovered for microwave polarizations that are far from circular by applying a static field at exactly along $z'$, which lies between the polarization ellipse's normal and semi-major axis.
Next, we investigate the robustness of this scheme to imperfections in the alignment of $\bm{\efield}$ and $z'$.
Their alignment can be achieved by controlling the static field \cite{covey2018}, or the microwave polarization \cite{bowers2015}, whichever is more practical.
Figure~\ref{fig:xy} shows the total loss rate due to both RSR and MIL as a function of undesired $x'$ and $y'$ components of the $\efield$ field.
Figure~\ref{fig:xy}(a) shows loss rates for $4\xi/\pi=3/4$,
which corresponds to 25~\% $\sigma_+$ circular and 75~\% $\sigma_x$ linear polarization.
Even though the polarization ellipse is very eccentric and closer to linear than it is to circular,
shielding of losses to below $10^{-12}$~cm$^3$/s can be achieved and requires alignment of the static field $\bm{\efield}$ and the $z'$-direction only to within 5$^\circ$.
Orientation of a static $\efield$ field to this precision is feasible \cite{covey2018}.
Figure~\ref{fig:xy}(b) shows loss rates for $4\xi/\pi=1/2$,
which is half way between circular and linear polarization and still too far from circular to realize shielding without the additional static field proposed here.
In this case, the tolerances on alignment of $\efield$ and $z'$ are even more forgiving, exceeding 10$^\circ$ for shielding of losses to below $10^{-12}$~cm$^3$/s.

In conclusion, we have proposed a modified scheme for microwave shielding that is robust against large imperfections in the circular polarization,
which is otherwise the main technical challenge for the experimental realization of microwave shielding.
The main idea is that polarization imperfections can generally be regarded as a spurious linear polarization component perpendicular to a perfectly circular polarization.
Application of a static field tunes the $m_n'=0$ component, addressed by the spurious linear polarization component, out of resonance with the microwaves.
The feasibility of the proposed scheme is illustrated by coupled-channels scattering calculations for bosonic RbCs molecules at 1~$\mu$K,
for which we recover effective shielding with achievable static field strengths, $\efield=1$~kV/cm, and microwave Rabi frequencies, $\Omega=10$~MHz,
even for polarizations far from circular.
Furthermore, the proposed scheme is robust against imperfections in the relative orientation of the polarization and static field.
The tolerance on their misalignment may exceed 10$^\circ$, depending on the eccentricity of the polarization.

Discussions with Jacob Covey, Zoe Yan, and Martin Zwierlein, regarding the experimental feasibility are gratefully acknowledged.
This work is supported by NWO Rubicon Grant No. 019.172EN.007 and the NSF through ITAMP.

\bibliography{Refs}
\end{document}